\documentclass[aps,prb,twocolumn,floatfix,%
showpacs,amssymb]{revtex4}

\usepackage{graphicx}

\begin{document}

\date{\today}

\title{Edge States and the Quantized Hall Effect in Graphene}

\author{Luis Brey$^1$ and H. A. Fertig$^2$}

\affiliation{1. Instituto de Ciencia de Materiales de Madrid (CSIC), Cantoblanco, 28049 Madrid, Spain\\
2. Department of Physics, Indiana University, Bloomington, IN 47405}

\begin{abstract}
We study edges states of graphene ribbons in the quantized Hall regime,
and show that they can be described within a continuum model (the Dirac equation)
when appropriate boundary conditions are adopted.  The two simplest
terminations, zigzag and armchair edges, are studied in detail.  For zigzag
edges, we find that the lowest Landau level states terminate in two types
of edge states, dispersionless and current-carrying surface states.  The latter
involve components on different sublattices that may be separated by distances
far greater than the magnetic length.  For armchair edges, the boundary conditions
are met by admixing states from different valleys, and we show that this leads
to a single set of edges states for the lowest Landau level and two sets for
all higher Landau levels.  In both cases, the resulting Hall conductance step
for the lowest Landau level is half that between higher Landau levels, as
observed in experiment.
\end{abstract}


\maketitle

\section{Introduction}

Recent progress in the processing of graphite have made possible the isolation
of two dimensional carbon sheets, known as graphene \cite{novoselov}.
This system has been studied theoretically for a number of years, because
when rolled up they form carbon nanotubes \cite{ando}.  The material
is unique because the underlying honeycomb lattice has a band structure
with Dirac points at the corners of the Brillouin zone, two of which are
inequivalent.  Undoped, the system has one electron per atom and the
Fermi energy surface passes directly through the Dirac points.  For low
energies and dopings the system may be described by the Dirac equation.

The stabilization of flat graphene sheets has allowed the
application of perpendicular magnetic fields and the observation of
the integer quantized Hall effect \cite{novoselov2,zhang}.  The
striking result in these experiments is the first step height in the
Hall conductance as a function filling factor, corresponding to
filling the lowest Landau level (with electrons or holes), is {\it
half} that of all subsequent steps. This behavior was expected
\cite{zheng} based on the bulk energy spectrum of graphene in a
magnetic field:  for a given spin species, there are pairs of Landau
level bands at positive energies, each with partners at negative
energies due to particle-hole symmetry.  The lowest Landau level
(LLL) has two levels precisely at zero energy,
each of which is its own particle-hole conjugate \cite{ando}.  This
property of the LLL results in its smaller
contribution to the Hall conductance.

The recent experiments on graphene studied ribbons that were relatively
narrow, with widths in the micron \cite{zhang} or submicron \cite{novoselov2}
range.  Under such circumstances, transport in the quantum Hall regime
is typically dominated by edge states \cite{edge_review}.  In this work
we study edge states for graphene ribbons in detail, focusing on the
simplest cases of a zigzag edge and an armchair edge.  We demonstrate
that a continuum description of edge states based on the Dirac equation is possible
with the adoption of appropriate boundary conditions.  In an edge state
description, the quantization of the Hall conductance is determined by
the number of edge state bands crossing the Fermi level.  The Hall
conductance results imply that the LLL supports only a single particle-like
and a single hole-like band (per spin) at each edge, while the higher
Landau levels have twice as many.  Our goal is to understand how and
why this happens, in a non-interacting picture.  (Because real spin plays
no role in this study, we will from here on assume that all the electrons
are spin-polarized, and refrain from explicitly noting the spin degree
of freedom in our discussions.)

We now summarize our results.  For zigzag edges, we show the correct
boundary condition is for the wavefunction to vanish on a single sublattice
across the edge.  In this case
the LLL supports two types of edge states, which we call current-carrying
and dispersionless surface states.  Both states have strong components at the boundary
of the system, but the former has equal weights on both types of sites
of the honeycomb lattice, whereas the latter exists essentially only on one
sublattice and has precisely zero energy.  Such zero energy surface states
are well-known to exist in graphene ribbons in the absence of a
magnetic field \cite{nakada,ryu}, and have been shown in tight-binding
calculations to persist when a field is applied \cite{castro}.
In the quantum Hall context we find that the dispersionless surface states
play a special role in forming two branches of edge states that do not
pass through the Fermi level for any non-zero doping.  The current-carrying
edge states are also remarkable in that the surface contribution on one
sublattice can be highly separated from a component well-inside the
bulk of the sample, on the other sublattice.  An interesting consequence
of this is that a particle injected near a zigzag edge should oscillate
back and forth between the edge and bulk, although presumably
such oscillations would be damped by many-body effects not
included in our study.

For armchair edges, we find the correct boundary condition is vanishing
of the wavefunction on both sublattices at the edge.  This is achieved
by mixing of wavefunctions from both Dirac points.
In this case, there are no dispersionless surface states,
and the LLL edge states behave differently than the higher
Landau levels for other reasons.
As we shall see,
the energetics of states from one of these valleys is generically higher
than from the other on a given sublattice, so that in the LLL
only one band of edge states can meet the boundary condition,
whereas in higher Landau levels there are two such bands.
The admixing of the two valleys leads to wavefunctions with
a characteristic spatial oscillation of the electron density with
wavevector equal to $\Delta K_x$, the difference between the
Dirac point wavevectors along the direction perpendicular to the edge.
Such oscillations should be observable in STM measurements.

\section{Preliminaries}
We begin by reviewing some generalities about electrons in graphene.
The lattice structure is a triangular lattice whose primitive
lattice vectors are ${\bf a} = a_0 (1,0) $ and ${\bf b} = a_0(1/2,
\sqrt{3}/2)$. There are two atoms per unit cell located at $(0,0)$
and at ${\bf d}= a_0(0,1/\sqrt{3})$. A simple tight-binding model
with only nearest neighbor hopping $t$ leads to a Hamiltonian with
Dirac points at the six corners of the Brillouin zone, only two of
which are inequivalent, and we take these to be ${\bf K}=\frac{2\pi}
{a_0} (\frac {1}{3}, \frac{1} {\sqrt{3}})$ and ${\bf
K}^{\prime}=\frac{2\pi} {a_0} (-\frac {1}{3}, \frac{1} {\sqrt{3}})$.
Wavefunctions can be expressed via
the ${\bf k} \cdot {\bf P}$ approximation \cite{ando,mele} in terms
of envelope functions $[\psi_A({\bf r}),\psi_B({\bf r})]$ and
$[\psi_A^{\prime}({\bf r}),\psi_B^{\prime}({\bf r})]$ for states
near the ${\bf K}$ and ${\bf K}^{\prime}$ points, respectively.
These can be conveniently combined into a 4-vector
$\Psi=(\psi_A,\psi_B,-\psi_A^{\prime},-\psi_B^{\prime})$. (The
reason for this sign convention will become apparent when we discuss
the armchair edge.) This satisfies a Dirac equation $H\Psi =
\varepsilon \Psi$,
\begin{equation}
H=\gamma a_0\,
\left( \begin{array} {cccc} 0 & -k_x+ik_y & 0 & 0 \\
-k_x-ik_y & 0 & 0 & 0 \\
0 & 0& 0 & k_x+ik_y \\
0 & 0 & k_x-ik_y & 0 \
\end{array} \right) \, \, \, , \label{hamilt_kp}
\end{equation}
with $\gamma=\sqrt{3}t/2$.  Note that ${\bf k}$ denotes the separation
in reciprocal space of the wavefunction
from the ${\bf K}$ (${\bf K}^{\prime}$) point
in the upper left (lower right) block
of the Hamiltonian.  To apply this Hamiltonian in the
presence of a magnetic field, one makes \cite{ando}
the Peierls substitution ${\bf k} \rightarrow -i {\bf \nabla} + e{\bf A}/c$
where ${\bf A}$ is the vector potential.

Before applying this procedure to systems with an edge, we point out
some interesting and useful properties of $H$.  Firstly, $H$ (and
the more exact tight-binding Hamiltonian from which it descended)
has \cite{ryu} chiral (i.e., particle-hole) symmetry, $\Gamma H
\Gamma=-H$, where $\Gamma= \left(\begin{array}{cc} \sigma_z & 0 \\ 0
& \sigma_z \ \end{array}\right)$, and $\sigma_z$ is the Pauli
matrix.  This tells us that a solution to the Dirac equation $\Psi$
with energy $\varepsilon$ has a particle-hole conjugate partner
$\Gamma\Psi$ with energy $-\varepsilon$.  Because of this, the
wavefunctions must be normalized on each sublattice separately:
$\int d{\bf r} [|\psi_{\mu}({\bf
r})|^2+|\psi_{\mu}^{\prime}({\bf r})|^2 ]= 1/2$, for $\mu=A,B$.
The solutions for states well away from the edge are well-known
\cite{ando}.  Taking ${\bf A}=-Bx\hat{y}$,
$\Psi=\frac{1}{\sqrt{L_y}} e^{ik_y y} \Phi $, and
$\psi_{\mu}(^{\prime})=\frac{1}{\sqrt{L_y}} e^{ik_y y}\varphi_{\mu}(^{\prime})$
with $L_y$ the
$\hat{y}$ extension of the sample, the wavefunctions retain their
valley index as a good quantum number, and the positive energy
wavefunctions may be written as
$\Phi=[\phi_{n-1}(x-(k_y+K_y)\ell^2),\phi_n(x-(k_y+K_y)\ell^2),0,0]$
for the ${\bf K}$ valley, and
$\Phi=[0,0,\phi_n(x-(k_y+K_y^{\prime})\ell^2),
-\phi_{n-1}(x-(k_y+K_y^{\prime})\ell^2)]$ for the ${\bf K}^{\prime}$
valley, with energies  $\varepsilon_n=\frac{\gamma
a_0}{\ell}\sqrt{2n}$. In these
expressions, $\phi_n$ is the $n$th harmonic oscillator state and
$\ell=\sqrt{eB/c}$ is the magnetic length.  The negative energy
states are easily obtained by reversing the signs of the
wavefunctions on the B sublattice.  For the case of the LLL, $n=0$,
and only one component of the 4-vectors for each valley is non-zero.
This means the particle-hole conjugate of these wavefunctions are
themselves. The bulk LLL wavefunctions do not have a clear particle-
or hole-like character.

\section{Zigzag Edge}

The geometry for a zigzag edge is illustrated on the top and bottom
edges of Fig. 1.  It is interesting to note that each atom at the
edge is of the same sublattice (say A). We shall see below that the
appropriate boundary condition is to set the wavefunction to zero on
a single sublattice (B), which we can understand to be the line of
lattice sites that would lie just above or below the system if the
bonds had not been cut to form the edge. 
In our discussions we will work with edges that lie along the $\hat{y}$
direction, so in what follows the coordinate axes in Fig. 1 will
be rotated by 90$^{\circ}$.
We begin by computing the band
structure for a tight binding model of a graphene ribbon with zigzag
edges, an example of which is illustrated in Fig. 2.  The flat
degenerate bands over a range of $k_y$ are Landau levels,
and, in the case of the LLL, dispersionless surface states
which we discuss below.   In a wide sample, there
is generically a large degeneracy within each Landau band, because
for the ${\bf K}~({\bf K}^{\prime})$ valley there are wavefunctions
peaked at $X_p=[k_y+K_y(^{\prime})+nG_y]\ell^2$, where $G_y$ is a
reciprocal lattice vector for the ribbon, and the integer $n$ can
take any value such that $X_p$ is between the sample edges. One may
conveniently reorganize the states by allowing all values of $k_y$
such that $-L_x/2 < [k_y+K_y(^{\prime})]\ell^2 < L_x/2$, with $L_x$ the ribbon
width, and assigning one state for each $k_y$ in the extended zone.

The prominent structure in Fig. 2 is the appearance of dispersing energy bands,
which occur when the wavefunctions approach the edges.  For the higher Landau
levels one observes two pairs of such bands, whereas for the LLL there is
only one such pair.  This means that a Fermi energy crossing
between the $n$th and $(n+1)$th  Landau levels yields a Hall conductance
$\sigma_{xy} = (2n+1)e^2/h$, as observed in experiment \cite{novoselov2,zhang}.

The unique behavior of the LLL edge states may be understood in terms of eigenstates
of the Dirac Hamiltonian with vanishing boundary conditions
on a single sublattice.  We begin by rotating the wavevectors in
Eq. \ref{hamilt_kp}, $k_x \rightarrow k_y,~k_y \rightarrow -k_x$
so that the zigzag edge lies along the $\hat{y}$,
and our wavefunctions then exist
in the space $x>0$.  Taking ${\bf A}=-Bx\hat{y}$ in this
coordinate system, and defining the ladder operator
$a=\frac{\ell}{\sqrt{2}}[-\tilde{k}_y+x/\ell^2+\partial_x]$,
with $\tilde{k}_y=k_y+K_y(^{\prime})$ for the ${\bf K}~({\bf K}^{\prime})$
valley,
the wavefunctions obey
\begin{equation}
{{2 \gamma^2 a_0^2} \over {\ell^2}} aa^{\dag}\varphi_A=\varepsilon^2 \varphi_A,
\quad \frac{2 \gamma^2 a_0^2}{\ell^2}a^{\dag}a\varphi_B=\varepsilon^2 \varphi_B
\label{sho}
\end{equation}
\begin{equation}
\frac{2 \gamma^2 a_0^2}{\ell^2}a^{\dag}a\varphi_A^{\prime}=\varepsilon^2 \varphi_A^{\prime},
\quad \frac{2 \gamma^2 a_0^2}{\ell^2}aa^{\dag}\varphi_B^{\prime}=\varepsilon^2 \varphi_B^{\prime}.
\label{shop}
\end{equation}
It is easy to see if one solves the equations for $\varphi_B$ and $\varphi_A^{\prime}$,
the remaining wavefunctions are determined by $\varphi_A= a\varphi_B/\varepsilon$
and $\varphi_B^{\prime}= -a\varphi_A^{\prime}/\varepsilon$.

For the zigzag edge, the boundary condition does not admix valleys,
and we can meet it for each type of wavefunction separately:
$\varphi_B(x=0)=\varphi_B^{\prime}(x=0)=0$.  Thus for the {\bf K} valley the spectrum
$\varepsilon^2_n(k_y)$ is identical to that of 
a quantum Hall edge with a sharp boundary \cite{halperin}.
The wavefunctions $\varphi_{B,n}$ similarly are identical
to their standard Hall edge counterparts, turning into states
in the $n$th Landau level as $k_y\ell^2$ moves well away from
the edge.  For $n \ge 1$, $\varphi_{A,n}$ is quite similar to
a state in the $(n-1)$ Landau level provided the center of the
wavefunction is not too close to the edge.

In the LLL ($n=0$),
$\varphi_A$ is qualitatively different.  Because LLL states are annihilated
by the ladder operator $a$, and $\varphi_B$ is similar to a bulk LLL state
when the center of the wavefunction is not too close to the edge,
$\varphi_A$ is extremely small, except close to $x=0$ where 
$\varphi_B$ vanishes and is forced to deviate from a bulk LLL state.
The result is that $\varphi_A$ is strongly confined to the surface,
and as the center of the $\varphi_B$ moves further into the interior
of the sample, $\varphi_A$ becomes increasingly so confined.
Despite this strongly localized form, the normalization of the wavefunctions
discussed in the previous section requires that fully half the probability
of finding the electron resides in this surface contribution.  We note that,
because $\varepsilon$ disperses with $k_y$, these surface states carry
current and contribute to the Hall conductivity.

Within the Dirac equation, the existence and form of these current-carrying
surface states can be examined with a variational approach.  We adopt a
trial wavefunction $\varphi_B(x) = w(x)\phi_0(x-k_y\ell^2)$ and require
$w(x \rightarrow 0)=0$.  One may easily confirm that $\tilde{\varepsilon}^2 =
2\int_0^\infty dx \bigl( \frac{dw}{dx} \bigr)^2 |\phi_0|^2$, where
$\tilde{\varepsilon}=(\ell/\sqrt{2}\gamma a_0) \varepsilon$.  A simple
choice for $w$ is $w=c_B(1-e^{-\lambda x})$, where $c_B$ is a normalization
constant and our variational parameter is $\lambda$.  With this choice
one finds $\varphi_A(x) \propto \exp\lbrace {-[x-(k_y-\lambda)\ell^2]^2/2\ell^2}\rbrace$,
so that if $\lambda>k_y$, $\varphi_A$ is confined to the surface.
The result of minimizing $\tilde{\varepsilon}^2$ is illustrated
in Fig. 3, where in the inset one sees that $\lambda$ increases 
faster than $k_y$, so that $\varphi_A$ becomes more
confined to the surface of the sample as $k_y$ increases and
$\varphi_B$ penetrates into the bulk.

Direct examination of wavefunctions from the tight-binding model
confirms this basic picture, except in one important respect.
Whereas the Dirac equation allows current carrying states with
$\varphi_A$ increasingly localized to the surface as $k_y$ grows
(and the peak position $X_p$ of $\varphi_B$ moves into the bulk),
the tight-binding results show that for $X_p >
\frac{1}{2}|K_y-K_y^{\prime}|\ell^2$  the LLL wavefunction on
sublattice A and the surface state on sublattice B appear as
separate states, and that the surface state now moves back into the
interior of the system with further increase in $k_y$. The
current-carrying surface state evolves into a dispersionless surface
state which we describe below, and the number of allowed
current-carrying surface states is limited.  This is clearly an
effect of the discreteness of the lattice that is not captured by
the Dirac equation: independent, highly localized surface states can
be written down for any $k_y$ in the continuum, but on the lattice
states at $k_y$ and $k_y+G_y$ have the same periodicity along the
$\hat{y}$ direction, and so cannot be made independent. Thus the
number of current-carrying surface states is limited to $\sim L_y
|K_y-K_y^{\prime}|/4\pi$. For any practical purpose, this may be
imposed by introducing a cutoff $k_c$ in the allowed $k_y$'s for the
current-carrying surface states.  For $k_y>k_c$, one may take the
${\bf K}$ valley wavefunction to have its bulk form.

The edge state wavefunctions of the ${\bf K}^{\prime}$ valley on
the A sublattice are analogously identical to the well-known
ones \cite{halperin}, although $\varepsilon_n^2$ is shifted
upward by a single unit ($2\gamma^2 a_0^2/\ell^2$) due to
the ordering of the operators in the last of Eqs. \ref{shop}.
For the LLL, the bulk states $\varphi_A^{\prime} \propto
\phi_0(x-X_p^{\prime})$, $\varphi^{\prime}_B=0$, with
$X_p^{\prime}=k_y\ell^2-K_y^{\prime}\ell^2$,
exactly satisfy Eqs. \ref{shop} with zero energy.  Remarkably, these
states are unaffected by the edge.
Moreover, because $K_y^{\prime} \ne
K_y$, there are values values of $k_y$ in the extended zone
for which $X_p^{\prime}<0$, and the state is confined to
the surface.  These are the dispersionless surface states:
they do not contribute to the Hall conductivity.
The LLL wavefunctions of the tight-binding
results around the center of the bands in Fig. 2(a) behave precisely as
the Dirac equation results suggest: they are strongly
confined to the surface, and continuously evolve into
bulk LLL states on the A sublattice as $k_y$ increases.
The dispersionless surface states supported by the
LLL are the reason that it carries only half the
Hall conductivity of the higher Landau levels
for the zigzag edge \cite{com}.

\section{Armchair Edge}

The armchair edge is illustrated
as the left and right edges Fig. 1, and the corresponding bandstructure
from a representative tight-binding calculation appears in Fig 2(b).
Here the edge runs along the $\hat{y}$ direction, and no rotation
of the figure is needed to represent our calculations.
Unlike the zigzag edge, the Landau bands all have dispersing states
in the same regions of $k_y$, but the LLL has one pair each of
hole-like and particle-like edge states, while all the higher
Landau levels have two.

To understand this from the viewpoint of the Dirac equation, we need
to impose appropriate boundary conditions.  In Fig. 1 one may see
that the termination consists of a line of A-B dimers, so it is
natural to have the wavefunction amplitude vanish on both
sublattices at $x=0$. To do this we must admix valleys, and require
$\varphi_B(x=0)=\varphi_B^{\prime}(x=0)$ and
$\varphi_A(x=0)=\varphi_A^{\prime}(x=0)$. 
Using the Dirac equation, and the fact that $K_y =
K_y^{\prime}$, this second condition
implies
$\partial_x\varphi_B|_{x=0}=-\partial_x\varphi_B^{\prime}|_{x=0}$.
To understand the effect of this on the solutions, it is convenient
to combine the $\varphi_B$'s into a single wavefunction defined for
$-\infty < x < \infty$: $\psi(x)=\varphi_B(x)\theta(x)+
\varphi_B^{\prime}(-x)\theta(-x)$, with $\theta(x)$ the step
function.  The boundary conditions then amount to $\psi(x)$ and its
derivative being continuous at $x=0$. (This was the reason for our
choice of relative sign in the 4-vectors of Section II.)  From Eqs.
\ref{sho} and \ref{shop}, it is easy to see that $\psi$ obeys a
Schroedinger equation $[-\partial_x^2
+U(x)]\psi(x)=\tilde{\varepsilon}^2\psi(x)$, with
$U(x)=\frac{\ell^2}{2}\big[(|x|/\ell^2-k_y)^2-1/\ell^2+(2/\ell^2)\theta(-x)\big]$.
For large $k_y\ell^2$, this double well potential, illustrated in
Fig. 4, has low energy states associated with the left well at
$\tilde{\varepsilon}^2 \approx 3/2,~5/2,\dots$, while for the right
well one has states at $\tilde{\varepsilon}^2 \approx
1/2,~3/2,~5/2,\dots$. We thus see there will be hybridization
leading to pairs of edge states for all the higher Landau levels,
whereas for the LLL there will be just a single such state.

The admixing of different valley states to meet the boundary
condition means that the wavefunction will oscillate with period
$2\pi/|K_x-K_x^{\prime}|$. The behavior can explicitly be seen in
Fig. 5, which illustrates a LLL edge state from the tight-binding
calculation. The apparent oscillation has precisely the period one
expects for the valley mixing we introduced in the Dirac equation to
meet the boundary condition.  Although the period of this
oscillation is very short (3.69\AA), 
it is in principle observable by STM measurements because the
samples can be open to their environment \cite{novoselov2}, in
contrast to GaAs systems.

\section{Conclusion}

In this paper we have studied the edge states of graphene
ribbons with zigzag and armchair terminations.  We found in both
these cases that a continuum description in the form of
the Dirac equation captures most features of the states
found in tight-binding calculations, provided the wavefunction
vanishes at the termination of the sample.  For zigzag edges,
we found the boundary condition can be met by wavefunctions
within a single valley, leading to two types of edge states
in the lowest Landau level, current-carrying surface states
and dispersionless surface states.  The latter of these
explains why the contribution to the quantized Hall coefficient
from the LLL is only half that of higher Landau levels.  For the
armchair edge, we found that admixing of valleys is necessary to
satisfy the boundary condition.  For higher Landau levels, there
are two hybridizations of the valley states for which this is
possible, whereas in the LLL there is only one.  This again
leads to a contribution to the Hall coefficient
from the LLL half the size of those from other occupied Landau levels.

{\bf Acknowledgements.} The authors thank F. Guinea and C. Tejedor for
useful discussions.  This work was supported by MAT2005-07369-C03-03
(Spain) (LB) and by the NSF through Grant No. DMR-0454699 (HAF).

\vfill\eject

\begin{figure}
 \vbox to 8.0cm {\vss\hbox to 10cm
 {\hss\
   {\includegraphics{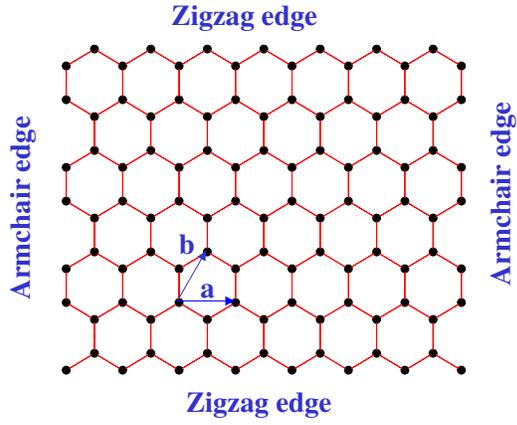}
   }
  \hss}
 }
\vspace{-20mm}
\caption{Ilustration of a graphene system with edges.  Top and bottom edges
are zigzag edges, left and right are armchair edges.
}
\label{fig1}
\end{figure}

\begin{figure}
 \vbox to 8.0cm {\vss\hbox to 10cm
 {\hss\
   {\includegraphics{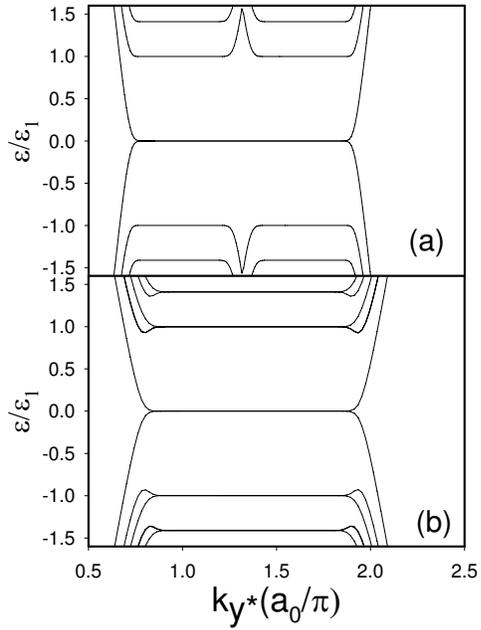}
   }
  \hss}
 }
\caption{
Examples of energy bands for a graphene ribbon with
periodic boundary conditions in the $\hat{y}$ direction and
edges in the $\hat{x}$ direction.  $B$=100T (0.00126 flux quanta per unit cell.)
Unit of energy $\varepsilon_1=\sqrt{2}\gamma a_0/\ell$.
(a) Ribbon with zigzag edges, 500 sites (530\AA) wide.
(b) Ribbon with armchair edges, 1000 sites (460\AA) wide.
}
\label{fig2}
\end{figure}

\begin{figure}
 \vbox to 5.0cm {\vss\hbox to 10cm
 {\hss\
   {\includegraphics{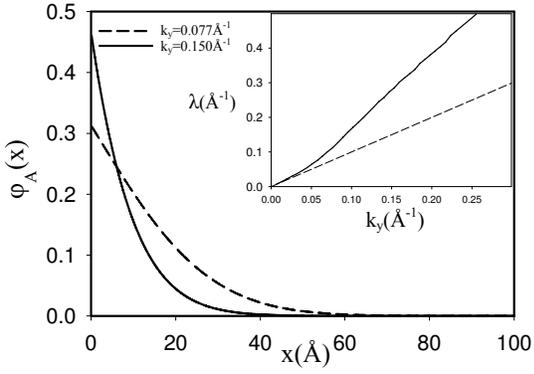}
   }
  \hss}
 }
\caption{
Surface part of current-carrying edge state for zigzag edge from variational
method described in text.  Main figure: $\varphi_A$ for two different choices
of $k_y$.  Inset: $\lambda_{min}$ (solid line) vs. $k_y$ (dashed line), demonstrating
that $\varphi_A$ becomes increasingly localized on the surface with
increasing $k_y$.
}
\label{fig3}
\end{figure}

\begin{figure}
 \vbox to 5.0cm {\vss\hbox to 10cm
 {\hss\
   {\includegraphics{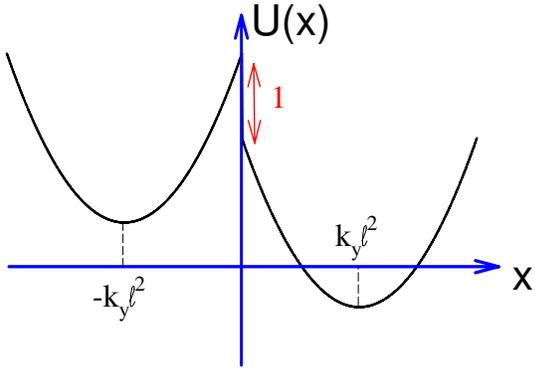}
   }
  \hss}
 }
\caption{
Potential $U(x)$ for an armchair edge.  See text.
}
\label{fig4}
\end{figure}

\begin{figure}
 \vbox to 5.0cm {\vss\hbox to 10cm
 {\hss\
   {\includegraphics{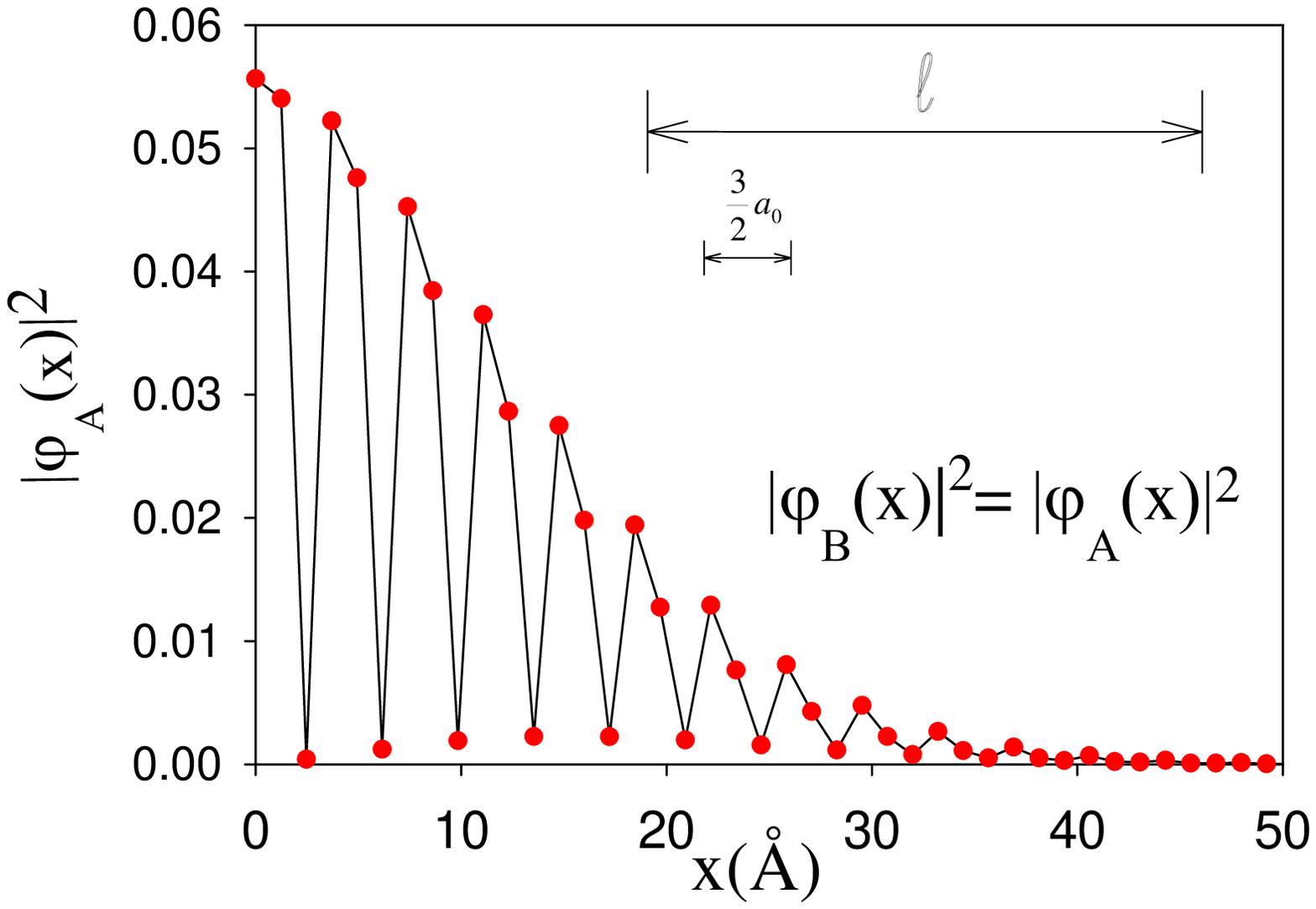}
   }
  \hss}
 }
\caption{
Squared wavefunction for an edge state of the armchair edge
from tight binding calculation.  $k_y=2.1\pi/a_0$
and $\varepsilon/\varepsilon_1=-0.202$.
Wavefunction penetrates sample over length scale $\ell$, while
oscillations due to valley mixing occur on a much smaller length
scale ($3a_0/2$).
}
\label{fig5}
\end{figure}


\begin{thebibliography}{99}
\bibitem{com} Terms that break the particle-hole symmetry 
(e.g., next nearest neighbor hopping) can introduce an energy dependence
in the dispersionless surface states.  However, 
the resulting edge states are not chiral and do not contribute
to the Hall conductivity.  See Ref. \onlinecite{castro}.
\end{thebibliography}
\end{document}